\documentclass{ws-ijmpd}
\usepackage[super,compress]{cite}
\def \D {\tilde{\nabla}}

\def\rd {\displaystyle{\cdot}}

\def \ts {\textstyle}
\def\om{\omega}

\def\3nab{\tilde{\nabla}}

\def\la {\langle}
\def\ra {\rangle}

\def\tl{\tilde}
\def\hsp5{\hspace{5mm}}
\newcommand{\sfrac}[2]{{\textstyle{#1\over#2}}}
\def\case#1/#2{\textstyle\frac{#1}{#2}}
\def\rd {\displaystyle{\cdot}}
\def\ts {\textstyle}
\def\ber {\begin{eqnarray}}
\def\eer {\end{eqnarray}}
\def\bea {\begin{eqnarray}}
\def\eea {\end{eqnarray}}

\def\ts {\textstyle}
\def\bc {\begin{center}}
\def\ec {\end{center}}
\def\case#1/#2{\frac{#1}{#2}}

\newcommand{\bw}{\begin{widetext}}
\newcommand{\ew}{\end{widetext}}
\newcommand{\nn}{\nonumber\\}

\newcommand{\be}{\begin{equation}}
\newcommand{\bse}{\begin{subequation}}
\newcommand{\ese}{\end{subequation}}
\newcommand{\ee}{\end{equation}}
\newcommand{\eei}{\end{eqnarray}\indent\indent}
\newcommand{\ba}{\begin{array}}
\newcommand{\ea}{\end{array}}
\newcommand{\bal}{\begin{eqnarray}}
\newcommand{\eal}{\end{eqnarray}}

\newcommand{\hs}{\,-\,}

\newcommand{\vP}{\varPhi}

\def\case#1/#2{\textstyle\frac{#1}{#2} }
\newcommand{\nb}{\nabla}

\newcommand{\gd}{g_{ab}}
\newcommand{\bver}{\begin{verbatim}}
\newcommand{\ever}{\end{verbatim}}

\begin{document}

\markboth{Amare Abebe,  Peter K.S. Dunsby and Deon Solomons}
{Integrability Conditions of Quasi-Newtonian Cosmologies in Modified Gravity}

%
\catchline{}{}{}{}{}
%

\title{INTEGRABILITY CONDITIONS OF QUASI-NEWTONIAN COSMOLOGIES IN MODIFIED GRAVITY}

\author{AMARE ABEBE}

\address{Department of Physics, North-West University,  Mafikeng 2735, South Africa\\
amare.abbebe@gmail.com}

\author{PETER K.S. DUNSBY}

\address{Department of Mathematics and Applied Mathematics, University of Cape Town, South Africa, Rondebosch 7701, Cape Town, South Africa\\
South African Astronomical Observatory,  Observatory 7925, Cape Town, South Africa\\
peter.dunsby@uct.ac.za}

\author{DEON SOLOMONS}
\address{Department of Mathematics and Applied Mathematics, University of the Western Cape, Bellville 7535, Cape Town, South Africa\\
deonsolomons@uwc.ac.za}
\maketitle


\begin{abstract}
We investigate  the integrability conditions of a class of shear-free perfect-fluid cosmological models within the framework of anisotropic fluid sources, applying our results to $f(R)$ dark energy models. Generalising earlier general relativistic results for time-like geodesics, we extend the potential and acceleration terms of the quasi-Newtonian formulation of integrable dust cosmological models about a linearized Friedmann-Lema\^itre-Robertson-Walker background and derive the equations that describe their dynamical evolutions. We show that in general, models with an anisotropic fluid source are not consistent, but because of the particular form the anisotropic stress $\pi_{ab}$ takes in $f(R)$ gravity, the general integrability conditions in this case are satisfied.
\end{abstract}

\keywords{cosmology; modified gravity; quasi-Newtonian.}

\ccode{PACS numbers:04.50.Kd, 04.25.Nx, 95.36.+x}


\section{Introduction}
The expansion, rotation and shear of time-like geodesic congruences are encoded in the differential properties of the geodesics, which in turn are related to the physical interpretation we give to the gravitational field equations. Shear and vorticity play important roles in the expansion dynamics of cosmological fluid models as described by the Raychaudhuri equation \cite{collins83, ellis2012R,herrera91, dadhich97, ellis2011S} and in the way distant matter can influence the local gravitational field.

There have been numerous studies on the role of shear in General Relativity, and  the special nature of shear-free cases in particular. In \cite{godel52} G\"{o}del showed  that  shear-free time-like geodesics  of some spatially homogeneous universes cannot expand and rotate simultaneously, a result later generalized by Ellis \cite{ellis67} to include inhomogeneous cases of shear-free time-like geodesics. Goldberg and Sachs \cite{goldberg62}, on the other hand, showed that shear-free null geodesic congruences {\it in  vacuo}  require an algebraically special Weyl tensor and this was generalized by Robinson and Schild \cite{robinson63} to include non-vanishing, but special forms of the Ricci tensor.

An interesting aspect of these shear-free solutions is that they do not hold in Newtonian gravitation theory \cite{ellis2011S, narlikar99, narlikar63}, even though Newtonian theory is a limiting case of General Relativity under special circumstances, namely at low-speed relative motion of matter with no gravito-magnetic effects (vanishing magnetic part of the Weyl tensor) and hence no gravitational waves. 

Recently, however, it has been shown \cite{Abebe2011} that there are some models of $f(R)$-gravity which exhibit Newtonian behaviour in the shear-free regime. For instance, some analytic $f(R)$ models \cite{Capozziello11} modify the Netwonian limit in the form of Yukawa correction to the gravitational field without requiring dark matter and enhance structure formation below the Compton wavelength. 

Cosmology with mimetic matter \cite{Sawicki13,Mirzagholi15} within the context of $f(R)$-gravity \cite{Chamseddine14} provides a more general $f(R)$ gravity theory in the sense that it is equivalent to a gravitational theory with two scalar fields, one of which accounts for the dark matter sector of the Universe, while the other is responsible for the dark energy content while satisfying the weak, null and dominant energy criteria and the Dolgov-Kawasaki instability criterion \cite{Shiravand16}.

In the Newtonian formulation of cosmological models, one considers potentials rather than forces in the dynamical evolution of spacetime, and a generalized concept of acceleration is introduced to represent the combined effects of gravitation and inertia \cite{ellis2012R}. Thus, Newtonian cosmologies are an extension of the Newtonian Theory (NT) of gravity and are usually referred to as ``quasi-Newtonian", rather than strictly Newtonian  formulations.

Despite there being no proper Newtonian limit for GR on cosmological scales, recent works on so-called {\it quasi-Newtonian} cosmologies \cite{elst98, maartens98, maartens1998} have shown that  gravitational physics (such as analysis of nonlinear collapse and structure formation using the Zel'Dovich approximation \cite{zeldovich70}) can be studied to a good approximation using such an approach. In \cite{elst98}, it was shown that non-linear quasi-Newtonian cosmologies are generally covariantly inconsistent in General Relativity. This inconsistency is due to the imposition of a shear-free and irrotational congruence condition, resulting in the vanishing of the magnetic part of the Weyl tensor \cite{trumper65}, thereby turning off the contribution from gravitational waves \cite{elst97}. This in turn puts strict constraints on the gravitational field \cite{maartens98,elst97, sopuerta97,kofman95}. On the other hand, when linearized around a Friedmann-Lema\^itre-Robertson-Walker (FLRW) model, these quasi-Newtonian models are consistent and provide a basis for the study of peculiar velocities in {\it almost FLRW} models \cite{maartens98,elst98}. It has been pointed out \cite{tsagas10} that peculiar motion can locally mimic the effects of dark matter in regions of a dust-dominated FRW universe endowed with bulk peculiar velocities.

Based on the ansatz for the evolution of the gravitational potential introduced by van Elst and Ellis \cite{elst98} and  Maartens' generalisation \cite{maartens98}, we first consider what happens if we modify the stress-energy tensor to include a a general anisotropic stress $\pi_{ab}$ and extend the general relativistic integrability conditions for the linearized models about the FLRW background. We show that while a general anisotropic source does not lead to integrability conditions that close on time propagation, $f(R)$ gravity is special because of the particular form $\pi_{ab}$ takes in this case. 

The paper is organized as follows:  in Section \ref{covsec} we give a brief summary of the covariant approach and the necessary covariant equations. In Section \ref{intsec}, we study the integrability conditions in a general fourth-order gravity and show in Sec \ref{frint}  that the 
linearized covariant equations are consistent by showing that the $f(R)$ covariant equations satisfy the integrability conditions.  
Finally in Section \ref{concsec} we discuss the results and give an outline of future work.

Unless otherwise specified, natural units ($\hbar=c=k_{B}=8\pi G=1$) will be used throughout this paper, and Latin indices run from 0 to 3. The symbol $\nabla$ represents the usual covariant derivative, we use the $(-+++)$ signature and the Riemann tensor is defined by
\begin{eqnarray}
R^{a}{}_{bcd}=W^a{}_{bd,c}-W^a{}_{bc,d}+ W^e{}_{bd}W^a{}_{ce}-
W^f{}_{bc}W^a{}_{df}\;,
\end{eqnarray}
where the $W^a{}_{bd}$ are the Christoffel symbols (i.e., symmetric in the lower indices), defined by
\begin{equation}
W^a{}_{bd}=\frac{1}{2}g^{ae}
\left(g_{be,d}+g_{ed,b}-g_{bd,e}\right)\;.
\end{equation}
The Ricci tensor is obtained by contracting the {\em first} and the
{\em third} indices
\begin{equation}\label{Ricci}
R_{ab}=g^{cd}R_{cadb}\;.
\end{equation}
The  action for $f(R)$ gravity can be written in these units as:
\begin{equation}\label{lagfR}
\mathcal{A}=\int {\rm d}^4 x \sqrt{-g}\left[\frac12 f(R)+{\cal L}_{m}\right]\;,
\end{equation}
where $R$ is the Ricci scalar, $f=f(R)$ is the general differentiable (at least $C^2$) function of the Ricci scalar and $\mathcal{L}_m$ corresponds to the matter Lagrangian. 

Finally, in a FLRW background universe, the non-trivial field equations resulting from the above action lead to the following equations governing the expansion history of the Universe \cite{carloni08, abebe2013}:
\begin{eqnarray}\label{raych}
&&\dot{\Theta}+\sfrac{1}{3}\Theta^2=-\frac{1}{2}\left(\mu+3p\right)\label{ray}\;,\\
&&\Theta^2= 3\mu+\frac{9K}{a^{2}}\;, \label{frid}
\end{eqnarray}
i.e., the {\em Raychaudhuri} and {\em Friedmann} equations. Here $\Theta$ is the expansion parameter, related to the Hubble parameter $H$ and the scale factor $a(t)$ via the standard relation $\Theta=3H=3\dot{a}/{a}\;,$
and the spatial curvature parameter $K$ takes values $\pm1 $ or $0$ depending on whether the FLRW models are  closed, open or  flat, respectively. One can also show that the Ricci scalar is given by
\begin{equation}
R=2\dot{\Theta}+\sfrac{4}{3}\Theta^2 \label{Riccci}+\frac{6K}{a^2}\;.
\end{equation}
\section{Covariant equations}\label{covsec}
For a given choice of the $4$-velocity vector field $u^a$, the Ehlers-Ellis covariant approach \cite{ehlers61, ellis71} gives rise to a set of fully covariant quantities and equations with 
transparent physical and geometrical meaning \cite{maartens98}.

The Einstein field equations can be written as
\be\label{efesmnm}
G_{ab}=T^{}_{ab}+T^{\ast}_{ab}=T_{ab}\;,
\ee
where $T_{ab}$ is the usual energy\hs momentum tensor (EMT) of standard matter given by
\be T_{ab}\equiv\frac{2}{\sqrt{-g}}\frac{\delta(\sqrt{-g}{\cal{L}}_{m})}{\delta \gd}\;,\ee
 and
$T^{\ast}_{ab}$ can be interpreted as the effective energy momentum tensor of any additional (imperfect) sources. In section 4 we will consider the case of $f(R)$ gravity. 
In this setting, the total EMT $T_{ab}$ of the entire cosmological medium, given by
\be
T_{ab}=\mu u_{a}u_{b}+ph_{ab}+q_{a}u_{b}+q_{b}u_{a}+\pi_{ab}\;,
\ee
is conserved, and so are the individual EMTs $T_{ab}$ and $T^{\ast}_{ab}$.

Here the projection tensor  $h_{ab}=g_{ab}+u_{a}u_{b}$ projects  the metric properties of the instantaneous rest spaces of observers \textit{orthogonal} to $u^{a}$ and the dynamical quantities
\ber
&&\mu=T_{ab}u^{a}u^{b}=\mu_{m}+\mu^\ast,~~~~~~~q_{a}=-T_{bc}u^{b}h^{c}_{a}=q^{m}_{a}+q^{\ast}_{a}\;,\nn
&&p=\sfrac{1}{3}(T_{ab}h^{ab})=p^{m}+p^{\ast},~~~~~\pi_{ab}=T_{cd}h^{c}{}_{\langle a}h^{d}{}_{b\rangle}=\pi^{m}_{ab}+\pi^{\ast}_{ab}\;,
\eer
are the total relativistic energy density, the relativistic momentum density (energy flux), the relativistic isotropic pressure and the trace-free anisotropic pressure of the total fluid. The effective energy density, isotropic pressure, heat flux and anisotropic pressure of standard matter and non-matter contributions are defined accordingly.

In such a treatment, the dynamics, kinematics and gravito-electromagnetics of the FLRW background is characterized 
respectively by
\ber
&&\D_a\mu=0=\D_a p\;, ~~~q_a=0\;,~~~\pi_{ab}=0\;,\\
&&\D_a\Theta=0\;,~~~A_a=0=\omega_a\;,~~~~\sigma_{ab}=0\;,\\
&&E_{ab}=0=H_{ab}\;.
\eer

The evolution and constraints of the above quantities are determined by applying the $1+3$-covariant decomposition  on the Bianchi and Ricci identities
\be\label{biricci}
\nb_{[a}R_{bc]d}{}^{e}=0\;,~~~~
(\nb_{a}\nb_{b}-\nb_{b}\nb_{a})u_{c}=R_{abc}{}^{d}u_{d}\;
\ee
for the total fluid 4-velocity $u^{a}$.  It is interesting to note here is that these dynamic and kinematic quantities  undergo transformations if a different $4$-velocity $\tl{u}^a$ is chosen. 

Because we are dealing with linearized perturbations of FLRW models, let us  choose the comoving (Lagrangian) $4$-velocity $\tl u^a$  to be of the linearized form \cite{elst98, maartens98}
\be
\tl {u}^{a} = u^{a}+v^{a}\;,~~~v_au^a=0\;,~~~v_av^a<<1\;,
\ee
where $v^{a}$  is a non-relativistic (``peculiar'') velocity that vanishes in the background. This is what is referred to as a {\it quasi-Newtonian} (Eulerian) frame  \cite{maartens98, batchelor00}. The matter pressure and vorticity both vanish in this frame,  while the matter energy flux $q^m_a$ is non-zero, {\it i.e.}, there is net particle flux due to the tilting of the  quasi-Newtonian frame relative to the comoving one. Moreover, both the isotropic and anisotropic pressures of matter vanish to linear order because they arise as second-order effects due to the relative motion\footnote{A generalized nonlinear transformation of quantities between the two frames is given in \cite{maartens98}.}:

\ber
&&\label{C1q}p_{m}=0\;, ~~~q^{m}_{a}=\mu_{m}v_{a}\;,~~~~\pi^{m}_{ab}=0\;,\\
&&\label{C2q}\om_{a}=0\;,~~~\sigma_{ab}=0\;.
\eer

To first  order, these quantities  evolve according to \cite{maartens98,  carloni08, abebe14}
\ber
&&\label{mueq}\dot{\mu}_{m}=-\mu_{m}\Theta-\tl\nb^{a}q^{m}_{a}\;,\\
&&\label{rayq}\dot{\Theta}=-\sfrac13 \Theta^2-\sfrac{1}{2}(\mu+3p)+\tl\nb_aA^a\;,\\
&&\label{qeq}\dot{q}^{m}_{a}=-\frac{4}{3}\Theta q^{m}_{a}-\mu_{m}A_{a}\;,\\
&&\label{edotq}\dot{E}^{\langle a b\rangle}=\eta^{cd\langle a}\tl\nb_cH^{\rangle b}_d-\Theta E^{ab}-\sfrac{1}{2}\dot{\pi}^{ab}
-\sfrac{1}{2}\tl\nb^{\langle a}q^{b\rangle}-\sfrac{1}{6}\Theta\pi^{ab}\;,
\eer
and are constrained by the following equations:
\ber
&&\label{R2q}C_0^{ a b}:=E^{a b}-\tl\nb^{\langle a}A^{b \rangle}-\sfrac{1}{2}\pi^{ab}=0\;,\\
&&\label{R4q}C_1^a:=\frac{2}{3}\tl\nb^a\Theta+q^{a}=0\;,\\
&&\label{propomq}C^2_a=\sfrac{1}{2}\eta^{abc}\tl\nb_bA_c=0\;,\\
&&\label{C3}C^3_{ab}:=E_{ab}-\frac{1}{2}\pi_{ab}-\tl\nb_{\la a}A_{b\ra}=0\;,\\
&&\label{R6q}C_4^{ a b}:=\eta^{cd\langle a}\tl\nb_cE^{\rangle b}_d-
\sfrac{1}{2}\eta^{cd\langle a}\tl\nb_c\pi^{\rangle b}_{d}\;,=0\;,\\
&&\label{B5q}C_5^a:=\tl\nb_bE^{a b}+\sfrac{1}{2}\tl\nb_{b}\pi^{ab}-\sfrac13\tl\nb^a\mu+
\sfrac13\Theta q^{a}=0\;.\\
\eer
The shear-free and irrotational condition \ref{C2q} and the gravito-electromagnetic (GEM) constraint \ref{R6q} result in the {\it ``silent" }\footnote{ In silent universes, the propagation equations decouple from the gradient, divergence and curl terms (the spatial derivatives, basically), thus forming ordinary differential evolution equations.} constraint
\be\label{silent}
H_{ab}=0\;,
\ee
i.e., no gravitational waves, and in turn showing that $q^m_a$ (as well as $v_a$) is irrotational:
\be
\eta_{abc}\D^bv^c=0=\eta_{abc}\D^bq^c_m\;.
\ee
It follows that for a vanishing vorticity, there exists a velocity potential $\phi$ \cite{maartens98, bert96} such that
\be
v_a=\D_a\mathcal{\phi}\;.
\ee
\section{Integrability Conditions}\label{intsec}
A constraint equation $C^A=0$ is said to {\it evolve consistently} with the evolution equations \cite{ maartens98, maartens1998, maart97} if
\be
\dot{C}^A=F^A{}_BC^B+G^A{}_{Ba}\D^aC^B\;,
\ee
where $F$ and $G$ are quantities that depend on the kinematic, dynamic and GEM quantities but not their derivatives. It has been shown \cite{maartens98, elst97} that the non-linear models are generally inconsistent if the silent constraint \ref{silent} is imposed,  but that those linearized around the FLRW background are  consistent. Thus, in order to find integrability conditions for quasi-Newtonian cosmologies, it suffices to show, using the appropriate transformations between the quasi-Newtonian and comoving frames, that these cosmologies  form a subclass of the linearized silent models.

The following shows the  mapping of the linearized kinematic, dynamic and GEM quantities from the quasi-Newtonian frame ($u^{a}$)  to  the comoving frame ($\tl{u}^{a}$) \cite{maartens98, elst97, abebe2013}: 
\ber
&&\tl\Theta=\Theta+\D^{a}v_{a}\;,\\
&&\tl{A}_{a}=A_{a}+\dot{v}_{a}+\sfrac{1}{3}\Theta v_{a}\;,\\
&&\tl\om_{a}=\om_{a}-\sfrac{1}{2}\eta_{abc}\D^{b}v^{c}\;,\\
&&\tl\sigma_{ab}=\sigma_{ab}+\D_{\la a}v_{b\ra}\;,\\
&&\tl\mu=\mu\;, ~\tl p=p\;,~\tl\pi_{ab}=\pi_{ab}\;,~\tl q^{\ast}_{a}=q^{\ast}_{a}\;,\\
&&\tl q^{m}_{a}=q^{m}_{a}-(\mu_{m}+p_{m})v_{a}\;,\\
&&\label{C7q}\tl E_{ab}=E_{ab}\;,~~\tl H_{ab}=H_{ab}\;.
\eer

Using Eqs. \ref{C1q}-\ref{C7q} one can covariantly describe linearized silent universe models by the following equations:
\ber
&&\tl p=0\;,~~ \tl q^{m}_{a}=0\;,~~\tl\pi^{m}_{ab}=0\;,\\
&&\label{C7c}\tl A_{a}=0\;,~~\om_{a}=0\;,~~\tl\sigma_{ab}=\D_{\la a}v_{b\ra}\;,\\
&& \tl E_{ab}=E_{ab}\;, ~~ \tl H_{ab}=0\;.
\eer
As we will see shortly, the special restriction placed on the shear in Eq. \ref{C7c} results in the integrability conditions for quasi-Newtonian models \cite{maartens98}.

We now see that for shear-free dust spacetimes (of which quasi-Newtonian models are a subclass) we can rewrite Eq. \ref{C3}  in the quasi-Newtonian frame as
\be
\label{C8}{\cal{E}}_{ab}\equiv E_{ab}-\sfrac{1}{2}\pi^{\ast}_{ab}-\tl\nb_{\la a}A_{b\ra}=0\;.
\ee
From Eq. \ref{propomq} and using the identity for a scalar $\varPhi$:
\be
\eta^{abc}\D_b \D_c\varPhi=-2\dot{\vP}\om_a\;,
\ee
it is clear that for an irrotational model,
\be
\eta^{abc}\D_b A_c=0\implies A_a\equiv\D_a\varPhi\;.
\ee
Here the scalar $\varPhi$ is the (peculiar gravitational) acceleration potential and corresponds to the covariant relativistic generalisation of the Newtonian potential. 
\subsection{First integrability condition}\label{fintsec}
Differentiating equation \ref{C8} with respect to cosmic time $t$, and using equations \ref{edotq} and \ref{R4q}, one obtains
\be\label{fic00} 
\tl\nb_{ < a}\tl\nb_{b>}\left(\dot{\vP}+\frac{1}{3}\Theta\right)+\left(\dot{\vP}+\frac{1}{3}\Theta\right)\tl\nb_{ < a}\tl\nb_{b>}\varPhi+\dot{\pi}^{\ast}_{ab}+\sfrac{2}{3}\Theta\pi^{\ast}_{ab}=0\;,
\ee
where, for any scalar function $X$, the identity
\be
\left(\D_{\la a}\D_{b\ra}X\right)^{.}=\D_{\la a}\D_{b\ra}\dot{X}-\sfrac{2}{3}\Theta\D_{\la a}\D_{b\ra}X+\dot{X}\D_{\la a}\D_{b\ra}\varPhi
\ee
has been used. For scalar perturbations we can write the anisotropic pressure $\pi^{\ast}_{ab}$ in terms of a potential $\Psi$ \cite{ellisCTAP97, dunsbyCTAP98}
\be\label{anipres}
\pi^{\ast}_{ab}=\D_{\la a}\D_{b\ra}\Psi\;,
\ee
then we can rewrite \ref{fic00} as
\be\label{fic} 
\tl\nb_{ < a}\tl\nb_{b>}\left(\dot{\vP}+\frac{1}{3}\Theta+\dot{\Psi}\right)+\left(\dot{\vP}+\frac{1}{3}\Theta+\dot{\Psi}\right)\tl\nb_{ < a}\tl\nb_{b>}\varPhi=0\;.
\ee

Equation \ref{fic} is the {\it First Integrability Condition} (FIC) for quasi-Newtonian cosmologies 
and is a generalisation of the  one obtained in \cite{maartens98}. 
The modified van Elst-Ellis condition \cite{elst98, maartens98} for the acceleration potential is thus generalized to \cite{abebe2013}:
\be\label{veec}
\dot{\vP}+\frac{1}{3}\Theta=-\dot{\Psi}\;.
\ee
An important consequence of this condition is the evolution equation of the 4-acceleration $A_a$. We derive this evolution equation from the gradient of \ref{veec} by means of the scalar commutation relation
\be\label{commdot}
(\tl\nb_{a}X)^{.}=\tl\nb_{a}\dot{X}-\frac{1}{3}\Theta\tl\nb_{a}X+\dot{X}A_{a}
\ee
and finally by using the shear-free constraint \ref{R4q}
\be\label{shco}
q_{a}=q^{m}_{a}+q^{\ast}_{a}=\mu_{m}v_{a}+q^{\ast}_{a}=\frac{2}{3}\tl\nb_{a}\Theta\;,
\ee
we arrive at
\be
\dot{A}_{a}+\left(\sfrac{2}{3}\Theta+\dot{\Psi}\right) A_{a}+\D_{a}\dot{\Psi}+\sfrac{1}{2}\mu_{m}v_{a}+\sfrac{1}{2}q^{\ast}_{a}=0\;.
\ee

\subsection{Second integrability condition}\label{sintsec}
To check for the consistency of the constraint \ref{C8} on any spatial hypersurface of constant time $t$, let us take the divergence of ${\cal{E}}_{ab}$, using the identity for projected vectors $A_{a}$:
\be\label{sic001}
\D^{b}\D_{\la a}A_{b\ra}=\sfrac{1}{2}\D^{2}A_{a}+\sfrac{1}{6}\D_{a}(\D^{c}A_{c})+\sfrac{1}{3}\left(\mu-\frac{1}{3}\Theta^{2}\right)A_{a}\;,
\ee
which, after some simplification, becomes
\be\label{sic0}
\D^{b}\D_{\la a}\D_{b\ra}\varPhi=\sfrac{2}{3}\D_{a}(\D^{2}\varPhi)+\sfrac{2}{3}\left(\mu-\sfrac{1}{3}\Theta^{2}\right)\D_{a}\varPhi\;.
\ee
Using Eqs. \ref{R2q}, \ref{R4q} and \ref{B5q},  Eq. \ref{sic0} can be re-written as
\ber\label{sic}
&&\D_a\mu-\frac{2}{3}\Theta\D_a\Theta-2\D_a(\D^2\varPhi)-2\left(\mu-\frac{1}{3}\Theta^2\right)\D_a\varPhi-2\D_a(\D^2\Psi)\nn
&&~~~~~~-2\left(\mu-\frac{1}{3}\Theta^2\right)\D_a\Psi=0\;,
\eer
which is the {\it Second Integrability Condition} (SIC).

This result is, in general, independent of the FIC (as is also the case without an anisotropic stress \cite{maartens98}). 

By propagating the (modified) van  Elst-Ellis condition we obtain the covariant {\it modified Poisson equation}:
\be\label{poisson}
\tl\nb^{2}\varPhi=\frac{1}{2}\left(\mu+3p\right)-\left[3\left(\ddot{\varPhi}+\ddot{\Psi}\right)+\left(\dot{\vP}+\dot{\Psi}\right)\Theta\right]\;.
\ee
Using Eqs. \ref{veec} and \ref{shco}, we are able to express  the peculiar velocity as
\be\label{vel}
v_{a}=-\frac{1}{\mu_{m}}\left[2\tl\nb_{a}\dot{\vP}+2\tl\nb_{a}\dot{\Psi}+q^{\ast}_{a}\right]\;.
\ee
By virtue of Eqs. \ref{mueq} and \ref{qeq}, $v_{a}$ evolves according to 
 \be\label{ve1}
\dot{v}_{a}+\frac{1}{3}\Theta v_{a}=-A_{a}\;.
\ee
This result is identical with that in GR. The peculiar velocity $v_a$ and the 4-acceleration $A_a$ decouple from  each other when one evolves Eq. \ref{ve1} to get the second-order propagation equation of the peculiar velocity. 
 By using  the Friedmann  and Raychaudhuri Eqs. \ref{frid}, \ref{rayq} in Eq. \ref{ve1} above, we obtain
\ber\label{ve2}
&&\ddot{v}_{a}+\left(\Theta +\dot{\Psi}\right)\dot{v}_{a}+\left[\sfrac{1}{9}\Theta^{2}+\sfrac{1}{3}\dot{\Psi}\Theta-\sfrac{1}{2}\mu_{m}-\frac{1}{6}\left(\mu+3p\right)\right]v_{a}\nn
&&~~~-\D_a\dot{\Psi}-\frac{1}{2}q^{\ast}_{a}=0\;.\nn
\eer
The corresponding general relativistic equation \cite{maartens98} reads
\be
\ddot{v}_{a}+\Theta\dot{v}_{a}-\left[\sfrac{1}{9}\Theta^{2}+\frac{2K}{a^{2}}\right]v_{a}=0\;,
\ee
where we have used the fact that $\frac{K}{a^{2}}=\frac{1}{3}\left(\mu_{m}-\sfrac{1}{3}\Theta^{2}\right)$. 

Now let us substitute for the peculiar velocity in Eq. \ref{ve1} by using the expression given in Eq. \ref{vel} to obtain, to  linear order, 
\ber\label{eqsub1}
&&\tl\nb_{a}\ddot{\varPhi}+\Theta\tl\nb_{a}\dot{\vP}+\left(\ddot{\vP}+\ddot{\Psi}-\sfrac{1}{2}\mu_{m}\right)\D_a\varPhi+\D_a\ddot{\Psi}+\Theta\D_{a}\dot{\Psi}\nn
&&~~~+\sfrac{1}{2}\dot{q}^{\ast}_{a}+\sfrac{7}{6}\Theta q^{\ast}_{a}=0\;.\nn
\eer
Using Eqs. \ref{rayq} and  \ref{veec} one can show that the acceleration potential $\varPhi$ satisfies
\be\label{ddotvarphi1}
\ddot{\varPhi}=\sfrac{1}{9}\Theta^{2}+\sfrac{1}{6}\left(\mu+3p\right)-\sfrac{1}{3}\tl\nb^{2}\varPhi-\ddot{\Psi}\;.
\ee
We are then able to replace the  coefficient $\ddot{\varPhi}$  in Eq. \ref{eqsub1} with expression \ref{ddotvarphi1} to arrive at the (linearized) equation
\ber\label{mvdot}
&&\tl\nb_{a}\ddot{\varPhi}+\Theta\tl\nb_{a}\dot{\vP}+\left[\frac{1}{9}\Theta^{2}+\sfrac{1}{6}(\mu+3p)-\sfrac{1}{2}\mu_{m}\right]\D_a\varPhi+\D_a\ddot{\Psi}+\Theta\D_{a}\dot{\Psi}\nn
&&~~~+\sfrac{1}{2}\dot{q}^{\ast}_{a}+\sfrac{7}{6}\Theta q^{\ast}_{a}=0\;.\eer
Let us take  the gradient of the modified Poisson equation \ref{poisson} and use the modified van Elst-Ellis condition \ref{veec} to obtain the second dynamical equation for the (peculiar gravitational) acceleration potential $\varPhi$:
\ber\label{fic2}
&&\tl\nb_{a}(\nb^{2}\varPhi)=\sfrac{1}{2}\tl\nb_{a}\mu+\sfrac{3}{2}\D_{a}p-3\tl\nb_{a}\ddot{\varPhi}-\Theta\tl\nb_{a}\dot{\vP}-\left( \dot{\vP}+\dot{\Psi}\right)\D_a\Theta-3\tl\nb_{a}\ddot{\Psi}-\Theta\tl\nb_{a}\dot{\Psi}\;.\nn
\eer
Now if one uses \ref{fic2} in \ref{sic}, the SIC becomes
\ber\label{siccc}
&&6\tl\nb_{a}\ddot{\varPhi}+6\Theta\tl\nb_{a}\dot{\vP}-\left(2\mu-\frac{2}{3}\Theta^2 \right)\D_a\varPhi+6\tl\nb_{a}\ddot{\Psi}+6\Theta\tl\nb_{a}\dot{\Psi}-\left(2\mu-\frac{2}{3}\Theta^2 \right)\D_a\Psi\nn
&&~~~-2\tl\nb_{a}(\D^{2}\Psi)-3\D_{a}p
=0\;.\nn
\eer


The two integrability conditions above are generally independent of each other. However, in the following section, we will show that Eq. \ref{siccc} is identical with Eq. \ref{sic} by virtue of Eq. \ref{mvdot} in the appropriate limits of $f(R)$ gravity theories, thus showing that given the First Integrability Condition, the Second Integrability Condition is identically satisfied..
\section{Integrability conditions for  $f(R)$ gravity}\label{frint}
If we specialise to $f(R)$ gravitational models, then applying variational principles of the action given by Eq. \ref{lagfR} with respect to the metric $g_{ab}$ results in the generalized field equations \eqref{efesmnm} reducing to
\be
G_{ab}=T_{ab}+T^{R}_{ab}\;,
\ee
where \be  T^{R}_{ab}=\frac{1}{2}g_{ab}\left(f-Rf'\right)+\nb_{b}\nb_{a}f'-g_{ab}\nb_{c}\nb^{c}f' +\left(1-f'\right)G_{ab}
 \ee
 is the curvature contribution to the EMT. It corresponds to the $T^{\ast}_{ab}$ we saw in Eq. \ref{efesmnm} and  is covariantly conserved. In these models, the  linearized curvature components of the thermodynamical quantities are given by
\ber
&&\label{mur}\mu_{R}=T^{R}_{ab}u^{a}u^{b}=\frac{1}{2}(Rf'-f)-\Theta f'' \dot{R}+\frac{1}{3}(1-f')\Theta^2+f''\D^2R\;,\nn
&& p_{R}=\sfrac{1}{3}(T^{R}_{ab}h^{ab})=\frac{1}{6}\left(3f-Rf'-2R\right)+f''\ddot{R}+f'''\dot{R}^{2}+\frac{2}{3}\Theta f''\dot{R}+\frac{1}{9}(1-f')\Theta^2\nn
&&~~~~~~-\frac{2}{3}f''\D^2R \;,\nn
&&q^{R}_{a}=-T^{R}_{bc}u^{b}h^{c}_{a}=\frac{1}{3}f''\Theta \tilde{\nabla}_{a}R-f'''\dot{R}\tilde{\nabla}_{a}R -f''\tilde{\nabla}_{a}\dot{R} -\frac{2}{3}(1-f')\tl\nb_a\Theta\;,\nn
&&\label{pir}\pi^{R}_{ab}=T^{R}_{cd}h^{c}{}_{\langle a}h^{d}{}_{b\rangle}=f''\tilde{\nabla}_{\langle a}\tilde{\nabla}_{b\rangle}R \;,
\eer
and  the total cosmic medium is composed of standard matter and the {\it curvature fluid}, with the total thermodynamical quantities given  by
\be\label{totaltherm}
\mu\equiv\mu_{m}+\mu_{R}\;,~~~\;p\equiv p_{m}+p_{R}\;,~~~
q_{a}\equiv q^{m}_{a}+q^{R}_{a}\;,~~~\;\pi_{ab}\equiv\pi^{R}_{ab}\;.
\ee
Comparing Eqs. \ref{anipres} and \ref{pir}, we conclude that, to linear order, $f(R)$ models correspond to modified gravitational theories with  $\Psi=f'$.
For these models, Eqs. \ref{mvdot} and \ref{siccc} are given by
\ber\label{mvdott}
&&2 f'\tl\nb_{a}\ddot{\varPhi}+2f'\Theta\tl\nb_{a}\dot{\vP}-\frac{2}{3}\left(\mu_m-\frac{1}{3}\Theta^2 f'\right)\D_a\varPhi+2f''\dot{R}\tl\nb_{a}\dot{\vP}+f''\D_a\ddot{R}\nn
&&+\frac{1}{3}\left[f-Rf'+2\Theta f''\dot{R}\right]\D_a\varPhi+\left(2f'''\dot{R}+\frac{4}{3}\Theta f''-2\dot{R}\frac{f''^2}{f'}\right)\D_a\dot{R}\nn
&&+\left[\frac{7ff''}{6}-\frac{\mu_m f''}{3f'}+\frac{4}{3}\Theta\dot{R}f'''-2\dot{R}^2\frac{f'''f''}{f'}+2\dot{R}^2\frac{f''^3}{f'^2}+f^{(iv)}\dot{R}^2+f'''\ddot{R}-\frac{1}{3}Rf''\right.\nn
&&\left.-\frac{1}{3}\dot{R}\Theta\frac{f''^2}{f'}\right]\D_a R=0\;\nn
\eer
and
\ber\label{sicccc}
&&6f'\tl\nb_{a}\ddot{\varPhi}+6\left(f''\dot{R}+f'\Theta\right)\tl\nb_{a}\dot{\vP}-\left[2\mu_m-\frac{2}{3}\Theta^2 f'+Rf'-f-2\dot{R}\Theta f''\right]\D_a\varPhi\nn
&&+3f''\tl\nb_{a}\ddot{R}+\left(6f'''\dot{R}+4\Theta f''-6\dot{R}\frac{f''^2}{f'}\right)\tl\nb_{a}\dot{R}+\left[\frac{7ff''}{2f'}-\frac{f''}{f'}\mu_{m}-Rf''\right.\nn
&&\left.-\dot{R}\Theta\frac{f''^{2}}{f'}+3f^{(iv)}\dot{R}^{2}+4\dot{R}\Theta f'''-6\dot{R}^2\frac{f'''f''}{f'}+6\dot{R}^2\frac{f''^{3}}{f'^2}+3\ddot{R}f'''\right]\D_{a}R=0\;,\nn
\eer
which can easily be shown to be identical because of the exact form the thermodynamic quantities given by Eqs. \ref{pir} take in $f(R)$ gravity.
\section{Results and Discussion}\label{concsec}
We have demonstrated the existence of two generally independent integrability conditions for generic fluid models with an anisotropic stress based on a linearized covariant consistency analysis of dust universes in the shear-free hypersurfaces (longitudinal) gauge originally 
developed for pressure free matter. The First Integrability Condition (FIC) is a result of temporal consistency requirement for the field equations, whereas the Second Integrability Condition (SIC) arises as a result of demanding spatial consistency of the field equations. 

We applied the analysis to the case of $f(R)$ theories, for which case we showed  that if one uses the modified van Elst-Ellis condition of the FIC, the SIC is identically satisfied. The exact form of the anisotropic pressure in Eq. \ref{pir} plays the key role for this result to be found.

We also derived the evolution equations for the acceleration and peculiar velocity, which follow from the generalized van Elst-Ellis condition for the 
acceleration potential $\varPhi$. 

As in GR, the velocity perturbations are scale-independent,  however because of the existence of the extra scalar degree of freedom arising from the modification of GR,  matter density fluctuations depend on the scale of the perturbations.  
A careful analysis of the matter density perturbations provide insight into how modifications of the theory of gravity change the effect of peculiar velocities on large scale structure formation. This is left for subsequent investigations.

A natural generalisation of this work worth investigating is to consider the nonlinear case in order to determine whether there are classes of $f(R)$ theories that lead to a consistent set of integrability conditions. 
%

\appendix
\section{Useful covariant identities}
In the standard covariant description, we project onto surfaces orthogonal to the $4$-velocity of the fluid flow using the projection tensor $h_{ab}\equiv g_{ab}+u_{a}u_{b}$ and $\tl\nb_{a}=h^{b}_{a}\nb_{b}$ is the 
spatially totally projected covariant derivative operator orthogonal to $u^{a}$, the comoving 4-velocity of fundamental observers. The covariant convective and spatial covariant  derivatives on a scalar function $X$ are respectively given by
\be
\dot{X}=u_{a}\nb^{a}X, ~~~~~\tl \nb_{a}X=h_{a}{}^{b}\nb_{b}X\;.
\ee
The geometry of the flow lines is  determined by the kinematics of $u^{a}$:
\begin{eqnarray}
&\nb_{b}u_{a}=\tl \nb_{b}u_{b}-a_{a}u_{b}\;, \label{delua}\\ 
&\tl \nb_{b}u_{a}=\sfrac{1}{3}\Theta h_{ab}+\sigma_{ab}+\omega_{ab}\;.
\label{projdelua}
\end{eqnarray}
From (\ref{delua}) and (\ref{projdelua}) we obtain an  important equation relating our key kinematic quantities:
\be
\nb_{b}u_{a}=-u_{b}\dot{u}_{a}+\sfrac{1}{3}\Theta h_{ba}+\sigma_{ba}+\omega_{ba}\;.
\ee
The RHS of this equation contains the acceleration of the fluid flow $\dot{u}_{a}$, expansion $\Theta$, shear $\sigma_{ba}$ and vorticity $\omega_{ba}$.
\section{Useful differential identities}
The following identities have been used to obtain the results in this paper  \cite{maartens98, maartens97}
\begin{eqnarray}
\eta^{abc}\D_b \D_cf =-2\dot{f}\omega_a \,,
\label{a13} \\
\D^2\left(\D_af\right) =\D_a\left(\D^2f\right) 
+{\ts{2\over3}}\left(\mu-\frac{1}{3}\Theta^2\right)\D_a f+2\dot{f}\eta_{abc}\D^b\omega^c
\,, \label{a19}\\
\left(\D_af\right)^{\rd} = \D_a\dot{f}-\frac{1}{3}\Theta\D_af+\dot{f}A_a \,,
\label{a14}\\
\left(\D_aS_{b\cdots}\right)^{\rd} = \D_a\dot{S}_{b\cdots}
-\frac{1}{3}\Theta\D_aS_{b\cdots}\,,
\label{a15}\\
\left(\D^2 f\right)^{\rd} = \D^2\dot{f}-\frac{2}{3}\Theta\D^2 f 
+\dot{f}\D^a A_a \,,\label{a21}\\
\D_{[a}\D_{b]}V_c = 
{\ts{1\over3}}\left(\frac{1}{3}\Theta^2-\mu\right)V_{[a}h_{b]c}
\,, \label{a16}\\
\D_{[a}\D_{b]}S^{cd} = {\ts{2\over3}}
\left(\frac{1}{3}\Theta^2-\mu\right)
S_{[a}{}^{(c}h_{b]}{}^{d)} \,, \label{a17}\\
\D^a\left(\eta_{abc}\D^bV^c\right) = 0 \label{a20}\\
\D_b\left(\eta^{cd\la a}\D_c S^{b\ra}{}_d\right) = {\ts{1\over2}}\eta^{abc}\D_b \left(\D_d S^d{}_c\right)\,,
\label{a18}
\label{a23}
\end{eqnarray}
where the vectors and tensors vanish in the background,
$S_{ab}=S_{\la ab\ra}$, and all identities
except (\ref{a13}) are linearized. (Nonlinear identities can be found
in \cite{maart97,hve96,maar97}.)



\begin{thebibliography}{99}   
\bibitem{collins83} C. Collins and J. Wainwright, Physical Review D {\bf 27} (1983) 1209.
\bibitem{ellis2012R} G.F.R. Ellis, R. Maartens R and M.A. MacCallum,  {\it Relativistic cosmology} (Cambridge University Press, 2012).
\bibitem{herrera91} L. Herrera, J. Iba{\~n}ez  and A. Di Prisco, {\it General Relativity and Gravitation} {\bf 23} (1991) 431-454.
\bibitem{dadhich97} N. Dadhich and L. Patel,  {\it  General Relativity and Gravitation} {\bf 29} (1997) 179-183.
\bibitem{ellis2011S} G.F.R. Ellis, {\it General Relativity and Gravitation} {\bf 43} (2011) 3253-3268.
\bibitem{godel52} K. G{\"o}del K,  Rotating universes in general relativity theory, in {\it  Proceedings of the International
Congress of Mathematicians}, eds.  L.M. Graves et al. (Cambridge, Mass. 1952)  {\bf 1} 175.
\bibitem{ellis67} G.R. Ellis, {\it Journal of Mathematical Physics} {\bf 8} (1967) 1171.
\bibitem{goldberg62} J. Goldberg and R. Sachs, {\it Acta. Phys. Polon.} {\bf 22} (1962) 13-23.
\bibitem{robinson63} I. Robinson I and A. Schild, {\it Journal of Mathematical Physics} {\bf 4} (1963) 484.
\bibitem{narlikar99} J.V. Narlikar, {\it Spinning universes in Newtonian cosmology on Einstein's Path} (Springer, 1999) 319-327.
\bibitem{narlikar63} J. Narlikar,  {\it Monthly Notices of the Royal Astronomical Society} {\bf 126} (1963)  203.
\bibitem{Abebe2011} A. Abebe, R. Goswami and P.K.S. Dunsby, {\it Physical Review D} {\bf 84} (2011) 124027.
\bibitem{Capozziello11} S. Capozziello  and M. De Laurentis, {\it Phys.Rept.} {\bf 509} (2011) 167-321.
\bibitem{Sawicki13}I. Sawicki and A. Vikman, {\it Phys. Rev. D} {\bf 87} (2013)  067301.
\bibitem{Mirzagholi15} L. Mirzagholi  and A. Vikman, {\it JCAP} 1506 (2015) 06, 028.
\bibitem{Chamseddine14} A. H. Chamseddine, V. Mukhanov  and A. Vikman, {\it JCAP} 1406 (2014) 017.
\bibitem{Shiravand16} Z. Haghani, S. Shahidi, and M. Shiravand, gr-qc/1507.07726.
\bibitem{elst98} H. van Elst and G.F.R. Ellis, {\it Classical and Quantum Gravity} {\bf 15} (1998) 3545.
\bibitem{maartens98} R. Maartens, {\it Physical Review D} {\bf 58} (1998) 124006.
\bibitem{maartens1998} R. Maartens, W.M. Lesame and G.F.R. Ellis, {\it Classical and Quantum Gravity} {\bf 15} (1998) 1005.
\bibitem{zeldovich70} Y.B. Zel'Dovich,  {\it Astronomy and Astrophysics} {\bf 5} (1970) 84-89.
\bibitem{trumper65} M. Tr{\"u}mper, {\it Journal of Mathematical Physics} {\bf 6} (1965) 584.
\bibitem{elst97} H. van Elst, C. Uggla, W.M. Lesame, G.F.R. Ellis and R. Maartens, {\it Classical and Quantum Gravity}
{\bf 14} (1997) 1151.
\bibitem{sopuerta97} C.F. Sopuerta, {\it Physical Review D} {\bf 55} (1997) 5936.
\bibitem{kofman95} L. Kofman and Pogosyan, {\it The Astrophysical Journal} {\bf 442} (1995) 30-38.
\bibitem{tsagas10} C.G. Tsagas, {\it Monthly Notices of the Royal Astronomical Society} {\bf 405} (2010) 503-508.
\bibitem{batchelor00} G.K. Batchelor, {\it An introduction to fluid dynamics} (Cambridge University Press, 2000).
\bibitem{carloni08} S. Carloni, P.K.S. Dunsby and A. Troisi, {\it Physical Review D} {\bf 77} (2008) 024024.
\bibitem{abebe2013} A. Abebe, PhD thesis,  University of Cape Town, 2013.
\bibitem{ehlers61} J. Ehlers, {\it Mainz Akademie Wissenschaften Mathematisch Naturwissenschaftliche Klasse} {\bf 11} (1961)
792-837.
\bibitem{ellis71} G. F. R. Ellis, {\it General relativity and cosmology}  (Academic Press, New York , 1971)  p. 104.
\bibitem{abebe14} A. Abebe, {\it Classical and Quantum Gravity} {\bf 31} (2014) 115011.
\bibitem{bert96} E. Bertschinger, Cosmology and large scale structure, in {\it Proc. of the Les Houches Summer School, Session LX}, eds.
R.. Schaeffer,  et al.  (Netherlands: Elsevier, 1996).
\bibitem{maart97} R. Maartens, {Physical Review D} {\bf 55} (1997) 463.
\bibitem{maartens97} R. Maartens and J. Triginer, {\it Physical Review D} {\bf 56} (1997) 4640.
\bibitem{hve96} H. van Elst H 1996, PhD thesis, University of London, 1996.
\bibitem{maar97} R. Maartens, G.F.R. Ellis and S.T. Siklos,  {\it Classical and Quantum Gravity} {\bf 14} (1997)1927.
\bibitem{ellisCTAP97} G.F.R. Ellis, {\it Astrophysics and Space Science Library} (Springer, 1997) pp.53-74.
\bibitem{dunsbyCTAP98} G.F.R. Ellis and P.K.S. Dunsby, {\it Current Topics in Astrofundamental Physics: Primordial Cosmology}  (Springer, 1998) pp 3-33.
\end{thebibliography}
\end{document}